# Optical Poling Reveals Hidden Molecular Restructuring in Multimode Fibers, Unlocking Ultra-Efficient Third-Order Nonlinearities

**Maxime Jonard[1], Thomas Larqué[2], Dan-Esli Bouyou Bouyou[2], Tigran Mansuryan[1], Alessandro Tonello[1], C. Sandt[3], Jean-René Duclère[2], Julie Cornette[2], Yago Arosa[4], Marc Fabert[1], Claire Lefort[1], Vincent Couderc[1], Maggy Colas[2]**

[1] *Université de Limoges, XLIM, UMR CNRS 7252, 123 Avenue A. Thomas, 87060 Limoges, France*
[2] *Université de Limoges, IRCER, UMR CNRS 7215, 12 Rue Atlantis 87068 Limoges, France*
[3] *SOLEIL Synchrotron, Gif-sur-Yvette, France*
[4] *iMATUS, University of Santiago de Compostela, Praza do Obradoiro S/N, Santiago de Compostela, Coruña 15782, Spain.*

vincent.couderc@xlim.fr

**Optical poling is a well-established technique for inducing $\chi^{(2)}$ nonlinearity, yet its impact on silica's molecular structure remains unexplored. Here, we report the first direct observation of molecular restructuring in large-core graded-index multimode fibers (MMFs) induced by optical poling, transforming the silica tetrahedral ring network. Through coherent light beating, this process converts large rings of more than four $SiO_4$ tetrahedra into smaller ones, altering both linear and nonlinear optical susceptibilities. Contrary to the assumption that poling efficiency stems solely from charge displacement, we show that structural modifications dominate, leading to record enhancements in third-order nonlinear processes, including geometric parametric instabilities (GPIs) and Kerr self-cleaning, despite a low modification of the Kerr coefficient. High-energy poling acts as an *in situ* annealing process, dynamically modulating the refractive index for unprecedented spatiotemporal light control.**
**These findings provide fundamental insights into silica's molecular dynamics under intense optical fields and open avenues for ultra-efficient nonlinear optical devices, enabling next-generation fiber-based photonics for high-power lasers, broadband light generation, and all-optical signal processing.**

## Introduction

Light propagation in multimode fibers involves complex interactions among spatial modes, leading to intricate spatiotemporal and spectral dynamics [1–6]. At high intensities, nonlinear phenomena, such as four-wave mixing, self-phase modulation, and stimulated Raman scattering, dominate, enabling broadband frequency conversion and energy redistribution across modes. In moderate power regimes, these interactions evolve into collective effects coupling Kerr modulation, Raman responses, and solitonic dynamics [1–6]. A notable phenomenon in this context is spatial beam reshaping, including self-cleaning and dynamic mode restructuring [1, 2, 7, 8]. However, silica's inherent inversion symmetry precludes bulk quadratic nonlinearity ($\chi^{(2)}$).
This constraint was first overcome in 1986 through all-optical poling process, breaking inversion symmetry in single-mode fibers and enabling second-harmonic generation (SHG) [9–13]. Recently extended to multimode fibers, optical poling not only induces SHG but also profoundly alters spatial and spectral beam properties [14, 15]. Yet, the mechanism driving the quasi-permanent inscription of $\chi^{(2)}$ nonlinearity remains unclear. Competing hypotheses exist: Farries et al. proposed non-phase-matched SHG seeds a self-written, periodic pattern via FF-SH interference [16], while Stolen et al. suggested a photoinduced static electric field drives charge displacement, creating local birefringence and quasi-phase-matching for $\chi^{(2)}$ processes [11]. These displaced charges, driven by photogalvanic effects, can initiate photocurrents and photoionization [12, 17, 18].
The resulting DC electric field maps onto the glass structure, aligning polar molecules akin to a Gaussian charge distribution's spatial derivative [19]. However, non-dipolar charge distributions, unexplained by simple directional photocurrents, have also been observed [20, 21], hinting at unexplored charge dynamics. Optical poling was assumed to modify charge distributions without altering silica's bulk structure. Yet, chemically inert silica fibers consistently exhibit reproducible $\chi^{(2)}$ induction, challenging conventional assumptions about symmetry-breaking in centrosymmetric materials. This raises a fundamental question: How does optical poling reshape silica's molecular landscape to sustain quasi-permanent nonlinearities?

**Theoretical Considerations:**
Amorphous silica's structure comprises a 3D network of $SiO_4$ tetrahedra forming cyclic structures. Its Raman spectrum includes D1 and D2 modes (associated with 3- and 4-member rings) and a broad mode at ~440 $cm^{-1}$ (Si-O-Si angle deformation) [22]. While fictive temperature (TF) modifications impact D1 and D2 modes, $GeO_2$ doping increases structural flexibility by reducing 3- and 4-member rings [22]. Structural properties are closely tied to thermal history and preparation methods, influencing density, refractive index, and physical properties [23–28]. Fiber drawing introduces radial TF distributions [29], which (introduced by Tool [30]) marking the transition from molten to supercooled liquid states. Variations in TF alter molecular organization, density, and charge distribution, affecting linear susceptibilitie [31].

In this letter, we reveal a novel mechanism for SHG via optical poling, driven by structural modifications of $SiO_4$ tetrahedral rings, the building blocks of multimode silica fiber cores. This all-optical process induces a quasi-permanent, periodic refractive index modulation, simultaneously tailoring linear ($\chi^{(1)}$), quadratic ($\chi^{(2)}$), and cubic ($\chi^{(3)}$) nonlinear susceptibilities. Beyond χ(2) grating inscription, we report the first observation of enhanced Kerr-type Geometric Parametric Instabilities (GPIs) [32–34] and a novel spatial self-cleaning effect for both fundamental and second-harmonic beams.
From a molecular perspective, we show that fibers with ring structures composed of more than 4 tetrahedra are most efficient for breaking centrosymmetry and achieving quasi-permanent $\chi^{(2)}$ inscription. Our results demonstrate that optical poling dynamically reshapes local silica structure, altering $SiO_4$ tetrahedral distribution and the fiber's nonlinear optical response.
These findings challenge conventional views on second-order nonlinearity in silica glasses and underscore the critical, overlooked role of fiber drawing in optimizing nonlinear processes. Our work elucidates the microscopic origins of $\chi^{(2)}$ induction and paves the way for next-generation fiber-based nonlinear photonics, where structural engineering enables unprecedented control over light-matter interactions.

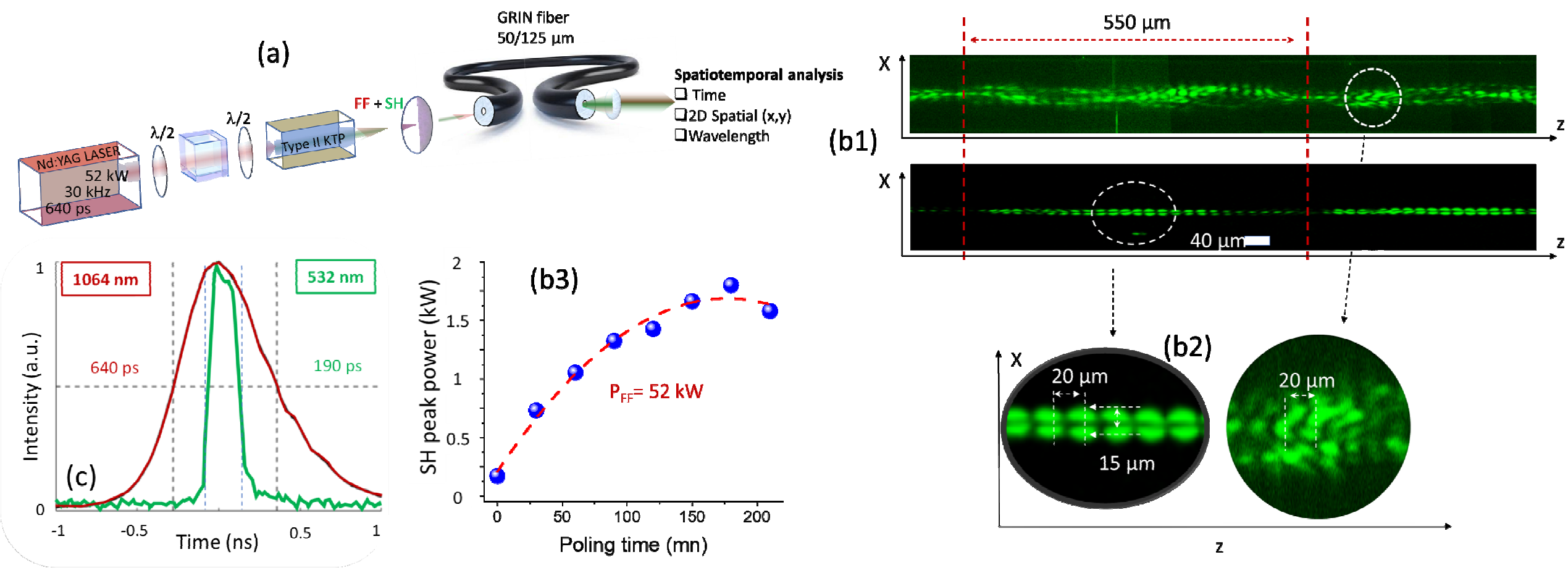

**Figure 1**: (a) Experimental setup of the optical poling process applied to Ge-doped multimode GRIN fibers. (b1–b3) SHG images of the $\chi^{(2)}$ nonlinearity induced by poling process. (b1) SHG images showing the $\chi^{(2)}$ nonlinearity distribution in the first (top) and last (bottom) 10 cm of the poled fiber (fiber length: 5 m; peak power: 52 kW). (b2) Magnified views of the images in (b1). (b3) Evolution of the SH output peak power versus poling time. (c) Temporal profiles of the fundamental and SH pulses after poling.

## Results

### Nonlinear SHG Process

The experimental setup (Figure 1(a)) uses a subnanosecond pulsed laser (640 ps, 1064 nm, 30 kHz repetition rate). Part of the infrared beam is converted to its second harmonic via a Type II KTP crystal. Both beams are coupled into a 5-meter-long 50/125 µm GRIN multimode fiber. After 30 seconds of exposure (FF: 45 kW, SH: 7 kW), only the infrared beam is coupled. This seeding process initiates $\chi^{(2)}$ evolution, and after over 200 minutes, a second harmonic peak power of over 1.75 kW is achieved (Figure 1(b3)).

Multiphoton microscopy reveals a complex $\chi^{(2)}$ structure in the fiber's first centimeters (Figure 1(b1), top), due to coherent beating of multimode FF and SH beams and before any spatial self-cleaning process. The beating period is reduced to 20 µm via optical rectification (Figure 1(b2)), and a 550 µm self-imaging period is observed despite multimode modulation. Kerr effect and self-imaging induce spatial self-cleaning [2-4] of the FF beam after tens of centimeters, transforming it into a quasi-single-mode bell-shaped beam (Figure 2(a)). This process organizes the beating into a structured transverse pattern (see figure 1(b1) bottom), primarily driven by the fundamental modes of FF and SH beams. The resulting $\chi^{(2)}$ pattern is then perfectly organized and shows two distinct longitudinal periodic evolutions.

Vertically, the poling dynamics create two birefringent areas separated by ~15 µm (Figure 1(b2)), due to charge displacement along the static electric field's polarization direction, generated by SH-FF beating [11]. The 2D transverse image (see Supplementary Material SM1) has been also measured by using SHG imaging microscope, showing circular pattern and explaining the two birefringent areas in Figure 1(b2).

Self-cleaning results in bell-shaped spatial profiles sat down on specular low energy beams for both FF and SH beams at the fiber output (Figure 2(a)) [2]. The SH wave's temporal profile narrows significantly from 640 ps to 190 ps (Figure 1(c)), exceeding typical pulse narrowing in birefringent crystals (factor of 1.4). Here, a compression factor over three is achieved, due to combined nonlinear spatiotemporal cleaning and frequency conversion.

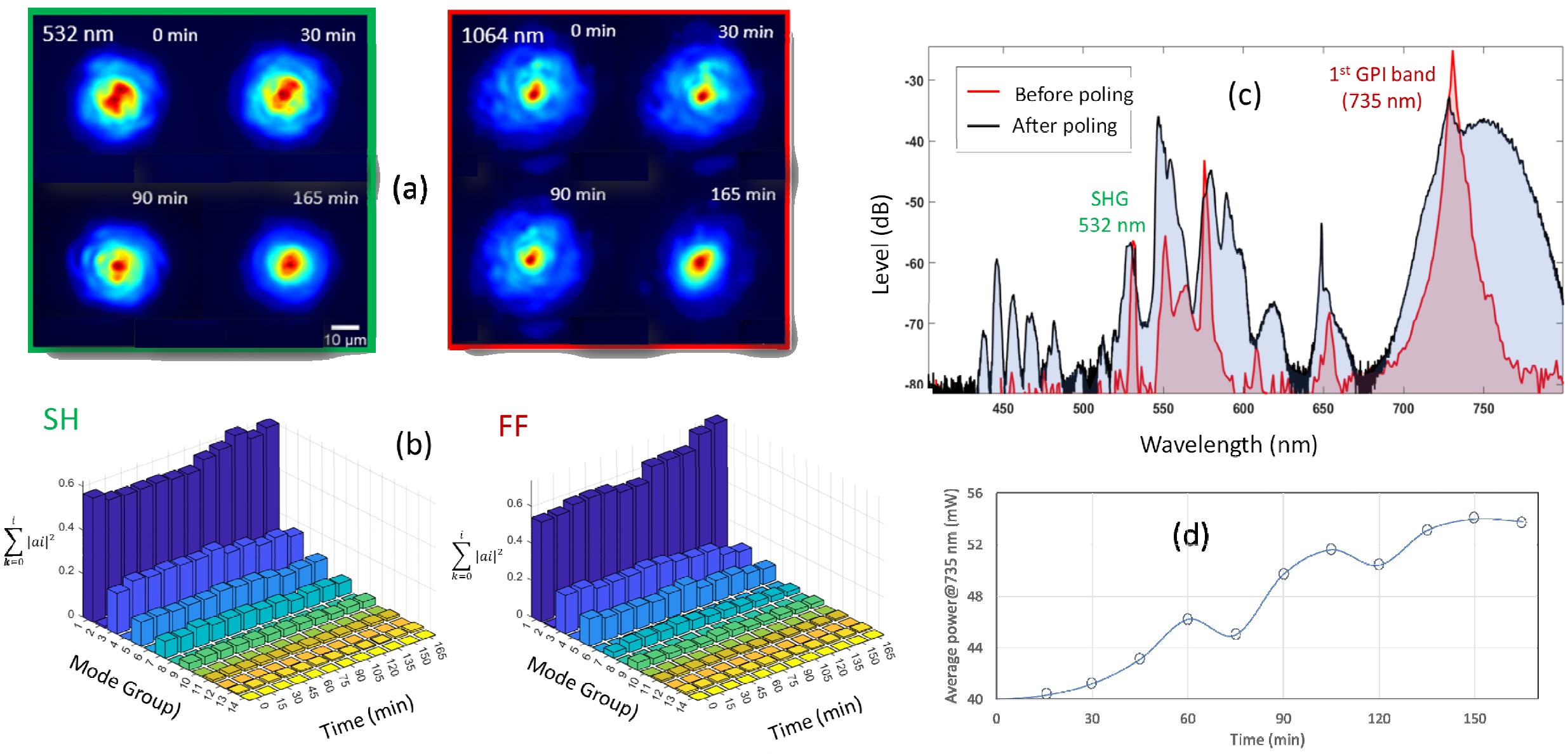

**Figure 2:** (a) Examples of near field images of the FF and SH output beams versus poling time; (b) Modal decomposition of the SH and FF output wavelengths plotted against poling time. (c) Output spectra in the visible domain, illustrating GPI line generation before and after a 165-minute of poling process. (d) Evolution of the output power of the first GPI line at 735 nm versus poling time.

## New Spatial Cleaning Driven by Poling Process

Beyond Kerr spatial self-cleaning [1–5], we observed additional evolution in the spatial profiles of both fundamental and SH beams as a function of $\chi^{(2)}$ writing time. This evolution, illustrated in Figures 2(a) and 2(b), reveals preferential energy redistribution toward the fundamental mode, reducing power propagating on higher-order modes. This suggests that poling process is sufficiently strong effect to introduce additional distortions in the core silica which acts as a booster effect for power exchange between modes, accelerating the spatial self-cleaning process.

## Improvement of GPI Nonlinear Process Efficiency

GPI, which exploits refractive index modulation due to self-imaging and the Kerr effect [34], can also benefit from a poling process that periodically alters the local refractive index of the fiber core. Furthermore, improvements in the self-cleaning process may also play a positive role in this four-wave mixing (GPI) process. To verify this hypothesis, we increased the fundamental beam power to excite both SH and GPI bands. GPI conversion efficiency is recorded during poling process by recording in time the output spectra (See Supplementary Material SM2). Figures 2(c) and 2(d) show the evolution of the 735 nm GPI anti-Stokes line power and spectral shape before and after poling. Thus, a significant improvement in the power carried by the first GPI line is clearly observed, with an increased spectral width indicating greater conversion. These observations could be due to structural modifications of the silica during propagation, reinforcing the $\chi^{(2)}$ nonlinearities but also moving the $\chi^{(1)}$ (for the phase-matching process) and $\chi(3)$ susceptibilities.

## Numerical Simulations

To understand poling-induced processes and nonlinear effects driven by $\chi^{(3)}$ coefficients, we performed simulations accounting for local linear refractive index modifications (see Methods and Supplementary Material SM6). Simulations reproduced SHG and confirmed dual modulation from FF-SH beating (20 µm) and transverse mode interactions (550 µm) (Figure 3(a)). Spatial evolution led to FF self-cleaning, extending to the SH beam (Figure 3(b)), increasing low-order mode energy at the expense of high-order modes (Figure 3(c)). This spatio-temporal modulation reinforces the random mode coupling as well as the disorder experienced by the FF which is at the origin of the self-cleaning efficiency [35]. GPI frequency conversion efficiency, governed by the 550 µm-period refractive index modulation, is also reproduced (Figure 3(d)), despite limited anti-Stokes broadening due to shorter propagation to avoid Raman effects.

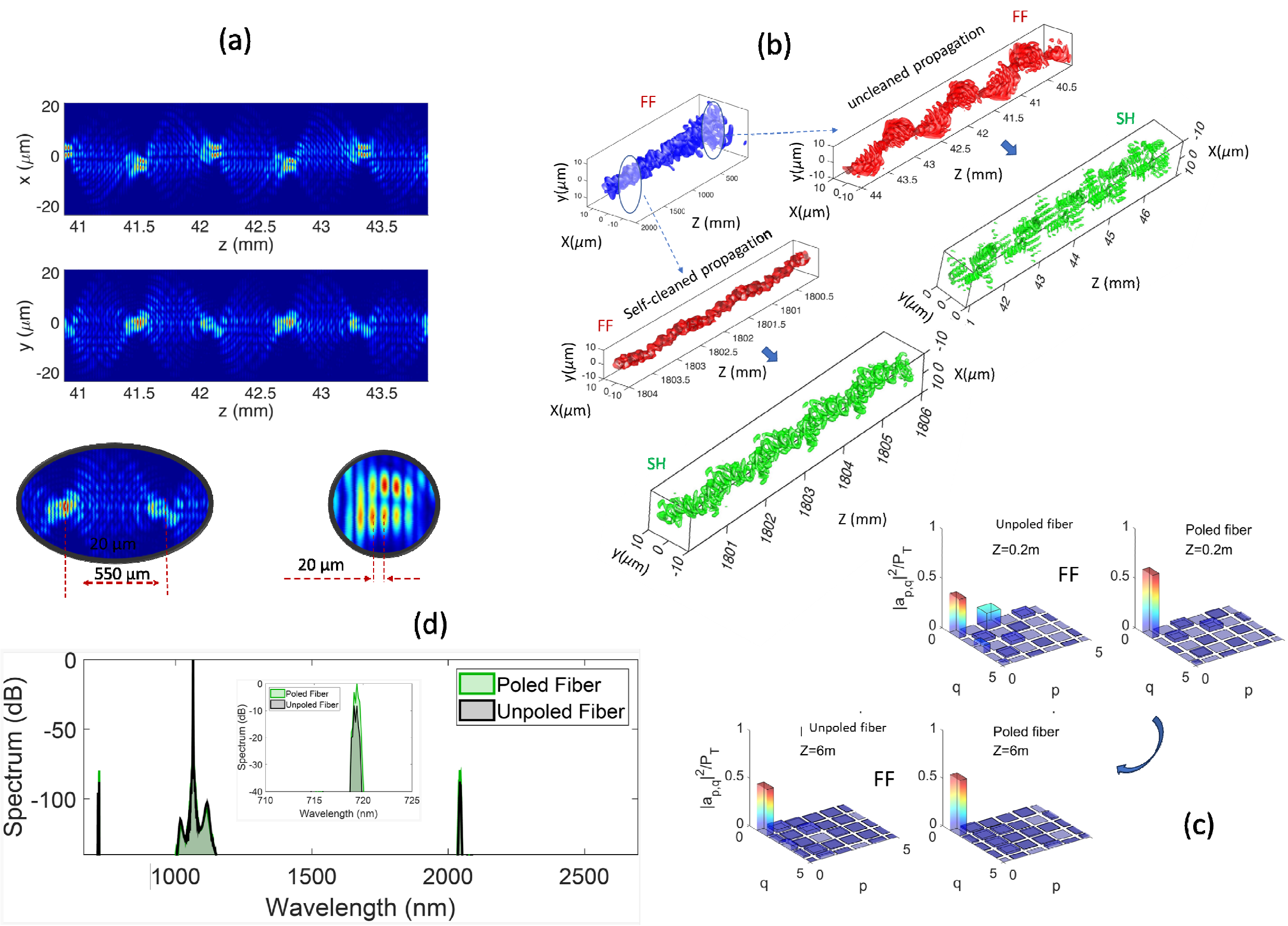

**Figure 3:** Numerical simulations of spatial and spectral evolution of the FF and SH beams: (a) 2D interference patterns between the FF and SH beams. (b) 3D multimode propagation of the FF and SH beams showing the self-cleaning process. (c) Modal decomposition of the FF beam for two different propagation lengths (0.2 m (top) and 6 m (bottom)), comparing scenarios without and with poling process (see methods and Supplementary Materials SM3). (d) Output spectra, with and without changes in the linear refractive index after the poling process, highlighting the enhanced efficiency of the GPI anti-Stokes line at 735 nm.

**Structural studies of raw silica core fibers:**

We investigated three Ge-doped multimode fibers (Fiber 1: 50/125 µm, Δn: 0.015, Alcatel; Fiber 2: 50/125 µm, Δn: 0.015, Thorlabs; Fiber 3: 62.5/125 µm, Δn: 0.028, Thorlabs) under identical poling conditions, yielding SH conversion efficiencies of 3.4%, 0.02%, and 0.96%, respectively. While the differences between Fibers 1 and 3 and Fiber 2 are expected due to their distinct core sizes and doping concentrations, the 170-fold disparity between Fibers 1 and 2, which share identical chemical and geometric compositions, is less intuitive (Supplementary Material SM4).

These differences suggest that the fiber drawing method influences the molecular arrangement, affecting the fibers' ability to undergo structural modifications under pulsed light. To explore this, we used optical photothermal infrared (O-PTIR) spectroscopy [36] and Raman analysis on the three fibers. O-PTIR spectroscopy revealed no significant differences in fictive temperature [37] between Fibers 1 and 2, indicating similar thermal histories (Supplementary Material SM5). Only Fiber 3, with its higher germanium concentration, showed distinct characteristics, evidenced by a reduction in the FWHM of the 1120 cm⁻¹ mode [23].

Raman maps and principal component analysis (PCA) of the fiber cores showed that Fiber 1 exhibited positive values for wavenumbers below 400 cm⁻¹, indicating a predominance of large $SiO_4$ tetrahedral rings, while Fiber 2 showed negative values for wavenumbers above 400 cm⁻¹, suggesting a higher proportion of small rings (Figures 4(a) and 4(b)). Fiber 3 displayed an intermediate situation (Figure 4(c)).

Raman map analysis of Fiber 1 before and after poling (Figures 4(d) and 4(e)) revealed that PC2 effectively discriminates between the pristine and poled states. Initially, the fiber exhibited a distribution dominated by large rings (below 425 cm⁻¹), whereas after poling, primary variations were observed in intermediate and small rings (above 550 cm⁻¹). These findings confirm that optical poling transforms the six-fold ring conformation into smaller ones (≤ 4 tetrahedra in a ring). Thus, these structural evolutions under poling process may modify $\chi^{(1)}$, $\chi^{(2)}$ and $\chi^{(3)}$ susceptibilities, which is commented in the discussion section.

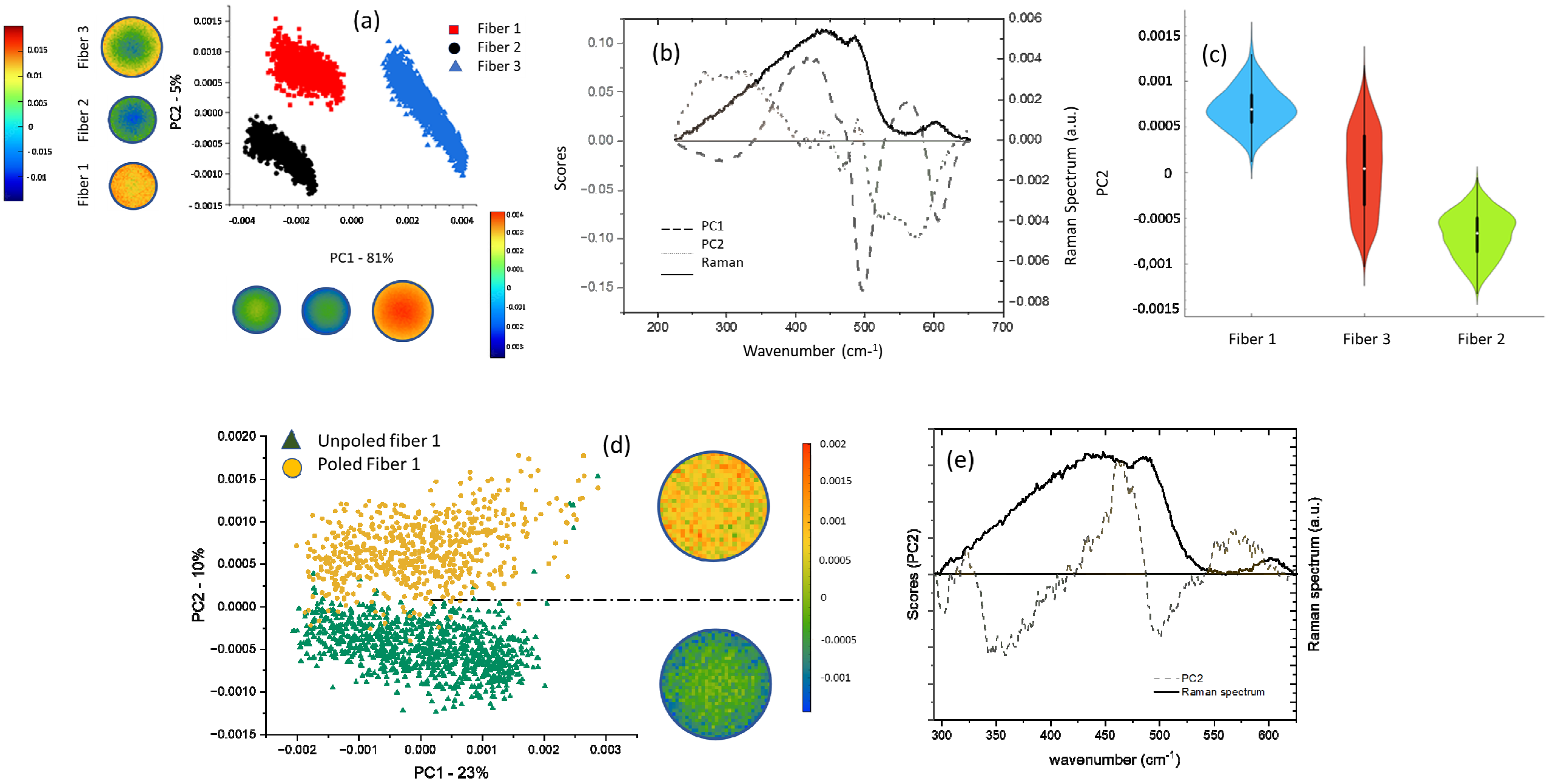

**Figure 4:** a) Projection of Raman data from the three fibers into the PC1–PC2 principal component space, alongside Raman maps of fiber scores reconstructed using the weights of PC1 and PC2, respectively, applied to the concatenated dataset. (b) PC1 and PC2 loadings plotted alongside a typical silica Raman spectrum for visual reference. (c) Violin plot representing the distribution of data projected onto PC2. (d) Projection of the data into the PC1–PC2 space, comparing conditions before and after the poling process for Fiber 1 showing the higher efficiency for SHG, with corresponding Raman maps reconstructed using the weights of PC1 and PC2. (e) PC2 loading plotted alongside a typical silica Raman spectrum for visual reference.

## Discussion

### a) Improving Third-Order Nonlinear Effects After Poling Process

Coherent nonlinear interactions between a weak SH beam and a strong FF beam in multimode GRIN fibers [14, 15] induce Bragg grating formation, generating two periodic modulations. The first arises from FF-SH beating, creating a static electric field that drives charge displacement and structural modifications, enhancing SH generation and coherence [20–21]. Initially dispersed due to intermodal interference, this distribution organizes under Kerr self-cleaning, converging toward a cleaned pattern mainly dictated by fundamental modes (Figures 1(b1) and 1(b2)).

The second modulation originates from intermodal FF-SH beating, generating a 550 μm-period nonlinear grating due to self-imaging and the Kerr effect [38, 39]. This results in energy transfer from high- to low-order modes, producing a spatially cleaned SH beam [2]. Our experiments achieved a 3.4% SH conversion efficiency with significant temporal compression (640 ps to 190 ps) which is attributed to cascaded nonlinear processes, including Kerr spatial self-cleaning and SH conversion [40, 41].

Surprisingly, beyond SH generation, we observed enhanced Kerr self-cleaning and GPI band generation versus poling time, relying on $\chi^{(1)}$, $\chi^{(3)}$ and 550 μm-period refractive index modulation [2, 34]. We hypothesized that poling may induce structural changes in the silica. Thus, we focused on the molecular structure of the fiber core, extending beyond the well-documented charge displacement mechanisms [20, 21] (see O-PTIR results in Supplementary Material SM5 and Raman spectroscopy data in Figure 4). Linear refractive index measurements by using the Arden Photonics nPA600 refractive profiler (Supplementary Material SM3) revealed a ~$6\times10^{-4}$ local distortion in the fiber core, demonstrating poling's direct impact on the $\chi^{(1)}$ susceptibility.

We then performed numerical simulations, incorporating an additional periodic modulation of the linear refractive index, synchronized with the self-imaging process, along the propagation axis. The supplementary refractive index modulation shows enhanced GPI band generation (Figure 3(d)), particularly at 735 nm and 1926 nm (Figures 2 and 3; Supplementary Material SM2). We also introduced this spatial modulation into simulations dedicated to the spatial self-cleaning process and we observed a supplementary power boost (≤15%) on the fundamental transverse mode (figure 3(c)) (see Supplementary Material SM6). Thus, an increase in the self-cleaning effect is clearly identified due to the 3D spatial optical poling process which reinforcing the synergistic interplay between spatial and spectral nonlinear dynamics in multimode fibers.

It is important to note that introducing higher disorder (Random mode coupling) into the simulations, supported by local modification of the poling process, did not lead to a direct spatial improvement in self-cleaning and even resulted in a decrease in its efficiency. This suggests that disorder, even when increased by the polling effect, appears to be a sensitive factor that can either enhance or hinder the spatial cleaning effect when it is too high.

### b) Fiber Core Structural Evolution After Poling Process

To further investigate the structural changes in the silica core, we replicated the poling experiment across multiple fibers. Our findings reveal that poling efficiency varies significantly not only between fibers with different chemical and geometrical characteristics but also between fibers with identical specifications manufactured by different suppliers. The most striking case involved two 50/125 µm graded-index (GRIN) optical fibers (see Supplementary Figure SM4). Despite their identical chemical and geometrical parameters, we observed a dramatic difference in poling efficiency, a factor of 170, with one fiber achieving a 1.75 kW peak power compared to just 10,3 W in the other. To elucidate this discrepancy, we conducted optical photothermal infrared (O-PTIR) spectroscopy and Raman spectroscopy analyses of the fiber cores, focusing on fictive temperature and molecular structure before and after the poling process.

O-PTIR spectroscopy and statistical analysis of data from three fibers (Supplementary figure SM4) demonstrated that only Fiber 3, which has a larger core diameter (62.5 µm vs. 50 µm) and a higher germanium doping concentration (8.1 at.% vs. 3.2 at.%), could be distinguished by these measurements (Supplementary Figures SM4 and SM5). This result aligns with literature findings and rules out fictive temperature as the source of the poling efficiency discrepancy between these fibers. Similar trends have been published in [42] and report higher silica densification while the Raman spectrum undergoes a shift as observed in our measurements as well as a local increase in the linear refractive index.

Through Raman spectroscopy, we successfully differentiated Fiber 1 from Fibers 2 and 3. While the first principal component (PC1), extracted from statistical computations, reproduced the O-PTIR results (i.e., distinguishing Fiber 3 due to its higher doping concentration), the second principal component (PC2) revealed a critical distinction between Fibers 1 and 2. The primary difference lies in the distribution of large tetrahedral rings (wavenumbers < 425 $cm^{-1}$), which is more pronounced in Fiber 1, exhibiting higher SH generation. These findings indicate that, despite their apparent similarity, the molecular structures of the two fibers differ significantly, particularly in the distribution of tetrahedral rings within the silica core. Consequently, Fiber 1's greater structural malleability enables it to undergo more substantial modifications when exposed to pulsed light during the poling process.
Theoretical models based on molecular hyperpolarizability and Kramers-Kronig relations link the molecular structure and chain length of materials to their overall nonlinear properties, as represented by the nonlinear refractive index ($n_2$) [43–45]. Generally, longer chain lengths correlate with higher nonlinear refractive indices due to increased electronic delocalization, though this relationship depends on the specific chemical and physical architecture of the material. In this context, Fiber 1 may exhibit a larger third-order nonlinearity, which amplifies the poling process efficiency, as observed in our experiments.

Beyond identifying the structural and molecular properties of Ge-doped GRIN silica fibers that enhance second-order nonlinearity writing efficiency, we also examined the structural evolution of these fibers post-poling. Our results, illustrated in Figures 4(d) and 4(e), reveal that the primary structural change in Fiber 1 after poling involves a reduction in the number of large 5- and 6-member tetrahedral rings and 3-member rings, accompanied by an increase in 4- and 2-member rings. This transformation suggests that the third-order nonlinearity after poling is not enhanced but rather constrained by the proliferation of smaller rings. Using the multimodal coherent anti-Stokes Raman scattering (M-CARS) method [46, 47], we detected a slight local modification of the third-order nonlinear susceptibility $\chi^{(3)}$ in the central region of the fiber core after poling (see Supplementary Document SM7).

Thus, optical poling is not solely controlled by charge delocalization, as previously thought, but is also profoundly influenced by the initial molecular structure, which is shaped by the fiber drawing process and modified by poling process.

## Methods

### Optical poling process

The optical poling process is performed using a 1064 nm Fourier-transform-limited pulse with a duration of 640 ps and a repetition rate of 30 kHz. The infrared beam features a single-mode Gaussian profile, which is spatially matched to the fiber core diameter (25 μm at full width at half-maximum intensity, FWHM) using a focusing lens. The maximum peak power coupled into the GRIN multimode optical fiber reaches 52 kW. A portion of the power is initially converted to the SH using a Type II potassium titanyl phosphate (KTP) crystal to initiate the poling effect. This seeding process is conducted within the fiber core with 7 kW of SH power and 52 kW at the fundamental wavelength for 30 seconds. Subsequently, only the fundamental infrared beam is maintained in the fiber by rotating a half-wave plate positioned before the KTP crystal. The poling time is extended up to 215 minutes. Spectro-spatial analysis of the output beam is performed using an optical spectrum analyzer (Ando) for spectral measurements and a CCD camera to record the spatial output profiles. The temporal shape of the pulses at both the FF and SH wavelengths is captured using an ultrafast oscilloscope with a 20 GHz bandwidth.

### O-PTIR spectroscopy

Infrared (IR) mapping was performed using Photothermal Spectroscopy Corp.'s mIRage infrared microscope, which detects the photothermal effect with a visible continuous-wave (CW) probe laser operating at 532 nm. Spectra were acquired in reflection mode, with a spectral data point spacing of 3 μm, using a 40× Schwarzschild objective featuring a numerical aperture of 0.78. The IR pump source covered a spectral range from 800 to 1400 $cm^{-1}$, with its power set to 100% (less than 3 mW), while the probe laser power was fixed at 25% (approximately 5.4 mW). The collected data were concatenated and preprocessed to enable direct comparison, thereby revealing correlations that are not apparent through conventional data processing methods.

### Raman spectroscopy

Raman mapping was carried out in point mode on a Renishaw Raman spectrometer at a wavelength of 785 nm with a 1200 line/mm grating and x100 objective. With an acquisition step of 1 micrometer, the hyperspectral data cube contains over 3,000 spectra. All vibrational datasets were analyzed using a multivariate approach based on principal component analysis similar to the ones of the OPTIR experiment [48].

### Numerical model of the poling distribution

We developed a first numerical model to describe the optical poling distribution along the fibre, emphasising its qualitative difference in distribution, from being spread out and irregular at the beginning to being more concentrated along the fibre axis at the end. In particular the experimental results show that optical poling develops around three different characteristic lengths: a short scale length (tens of micrometres) depending on the mismatch $\Delta k$ between the FF and SH beam; a second scale developing at the self-imaging period, intensifying the local equivalent quadratic susceptibility with a scale of half a millimetre; and a third, longest evolution scale over a metre of propagation, caused by the nonlinear beam reshaping of the FF, which also causes a remarkable difference in the distribution of the local quadratic susceptibility caused by the optical poling process.

We considered a GRIN fiber with radius R=26 μm, $n_0$=1.47, $n_{clad}$=1.457, Kerr coefficient $n_2=2.6\times10^{-20}\,m^2/W$.

Let us focus on the last two length-scales, the second and third. Both can be numerically simulated by solving the scalar Helmholtz equation for the fundamental mode, which includes the truncated parabolic transverse profile of the refractive index and the nonlinear Kerr effect. The equation for the complex envelope *A(x,y,z)* neglecting time-domain effects reads as:

$$-i\frac{\partial A}{\partial z} = \frac{\nabla_\perp^2 A}{2k_0} + \frac{\omega_0}{2n_0 c}[n(x,y,z)^2 - n_0^2]A + \gamma|A|^2 A \qquad (1)$$

where n(x,y,z) accounts for the truncated parabolic profile of the refractive index and $\gamma$ is the Kerr nonlinear coefficient. The disorder was implemented by weak random deformations of the circular refractive index shape, deforming to an ellipse with maximum variation between the larger and smaller axes of 0.2 μm, and random axis orientation. The deformation was applied with a coarse step of 5 mm average length and 10% variation to avoid unwanted resonances. The input beam diameter was of 30 μm FWHMI and an input tilt angle of 0.5° and the maximum peak power density was of 100 GW/cm$^2$.

When disorder is considered, the known effects of beam self-cleaning, repopulation of the fundamental mode, and spread of optical power to high-order modes are observed.

This effect can be also emphasized by computing the local intensity I(x,y,z) and by tracing the corresponding iso-intensity surfaces at 50% of the function $I/max(I)$ where the intensity is normalized to its local maximum along the propagation coordinate z. However, the beam shape reflects a somewhat cylindrical profile, whereas the analysis with the nonlinear microscope unveils that the poling zones are distributed in a sort of double-helix cylinder with a poor concentration of nonlinear susceptibility exactly along the fibre axis. To have a reasonable numerical model to move from the optical field distribution to the material deformation induced by the optical poling we referred to Ref [49].

Basing on these results we computed the iso-intensity surface of its local gradient normalized to its local maximum (see figure 3).

$$F = \frac{|\nabla_\perp I^2|^2}{\max(|\nabla_\perp I^2|^2)}$$

While the last length scale can be observed by integrating equation 1 and applying the procedure over a few metres of propagation, the second submillimetre length scale can be appreciated by storing for a limited segment of propagation more intermediate results. Note that the second harmonic generation has never been simulated and the present analysis is limited to consider it forced by a term proportional to $I^2$. To add to the picture the first and shortest scale, we simply multiplied the local intensity by $sin^2(\Delta kz)$. Finally, we can speculate that a function showing the three length scales over which nonlinear susceptibility develops can be visualised by computing a function of the form $FI^2 sin^2(\Delta kz)$.

## Numerical model of nonlinear propagation in a disordered and optically poled fibre

It is known that disorder may accelerate the cleaning process and we attempted to quantify the effect of poling [37]. Our experimental results indicate a weak but significant difference in beam cleaning between a poled fibre and an unpoled fibre. To evaluate the effect of optical poling-induced disorder on beam dynamics, we numerically applied two different linear refractive index perturbations along the fibre to reproduce the qualitatively different susceptibility distribution between the start and end of the fibre, as the local variations of the quadratic susceptibility correspond also to variations in the refractive index. We then integrated numerically eq.1 implementing a in a coarse step three different types of local deformations:

1) We constantly deformed the fibre core shape from circular to elliptical to cause localised linear mode coupling. We applied random deformations of the circular core, up to 0.18 μm in extension, random angles between the largest and smallest axes, and a coarse length with an average of 5 mm and a random variation of 10% to avoid longitudinal resonances.
2) For the first 60cm of numerical propagation we added within the same coarse step a local change in the refractive index of $\Delta n_{max} = 10^{-4}$ in a disc of 5 μm localized in the fibre core at a random distance bounded in a circular crown from 8 μm to 15 μm from the fibre axis. This effect would mimic the local linear refractive index variation caused by the optical poling effect and its random distribution, which only slightly affects the fibre axis for the first meter of propagation.
3) For the remaining propagation, we introduced a permanent variation in the transverse refractive index profile by adding a constant value $\Delta n_{max} = 10^{-4}$ to the truncated parabolic profile in a cylinder of radius 10 μm and centred on the fibre axis.

## Numerical model of GPI in poled fibers

The numerical simulation of the GPI for a GRIN fiber of core radius R and relative refractive index difference $\Delta$ requires the extension of the model to include explicitly the time domain *A(x,y,t,z)* and to account for the different weight of the diffraction at the different spectral components. For the temporal spectrum $\tilde{A}(x,y,\omega,z)$ we considered the following model [50].

$$\frac{\partial \tilde{A}}{\partial z} = \frac{i}{2k(\omega)}\nabla_{\perp}^{2}\tilde{A} + iD(\omega)\tilde{A} - ik(\omega)\Delta\frac{r^2}{R^2}\tilde{A} + i\gamma\mathcal{F}[(|A|^2)A]$$

Since the propagation distance was short, we omitted the coarse step for linear random mode coupling. The parabolic refractive index variation was implemented by considering the molar concentrations of $GeO_2$ and $SiO_2$ using the Fleming formula. This model was enhanced to account for the permanent modification of the refractive index in poled fibers. We hypothesised that a similar effect to the permanent local refractive index variation caused by poling at the fibre's self-imaging period could be modelled by introducing a periodic refractive index variation, similar to a long-period Bragg grating. In particular we considered the periodic introduction of local variation of the refractive index in the form of disks with radius $R_D$=6 μm and refractive index difference of $10^{-5}$ with respect to the core. These disks were aligned with the fibre axis and had a longitudinal thickness of 80 μm. The period of these disks of refractive index perturbation, was chosen to be equal to the self-imaging period for radially symmetric beams. For unpoled fibre this additional permanent longitudinal variation of the refractive index was removed.

**Acknowledgements**

VC thanks the National Research Agency under the Investments for the future programme with the reference *ANR-10-LABX-0074-01 Sigma-LIM, the ANR-23-CE08-0021-02, ANR-21-ESRE-0007 ADD4P, ANR-24-CE46-7295-01-ACT, and the EIC Pathfinder project "Multiscope" N°: 101185664. YA acknowledges receiving a postdoctoral fellowship (ED481D-2024-001) from the Xunta de Galicia (Spain).*

**Author contributions**

M. J., T. M., Y. A., V. C., J. C., C. L., D-E B. B., T. L., carried out the experiments. A. T., Y. A., M. C., J-R. D., C. S., realized numerical simulation and proposed theoretical interpretation. All authors analysed the obtained results, and participated in the discussions and in the writing of the manuscript.

**Additional information**

**Competing financial interests**

The authors declare no competing financial interests.

**Data availability**

Data underlying the results presented in this paper are not publicly available at this time but may be obtained from authors upon reasonable request.

**Code availability**

Codes used in this paper are not publicly available at this time but may be obtained from authors upon reasonable request.

# Supplementary material

# Optical Poling Reveals Hidden Molecular Restructuring in Multimode Fibers, Unlocking Ultra-Efficient Third-Order Nonlinearities

**Maxime Jonard[1], Thomas Larqué[2], Dan-Esli Bouyou Bouyou[2], Tigran Mansuryan[1], Alessandro Tonello[1], C. Sandt[3], Jean-René Duclère[2], Julie Cornette[2], Yago Arosa[4], Marc Fabert[1], Claire Lefort[1], Vincent Couderc[1], Maggy Colas[2]**

[1] *Université de Limoges, XLIM, UMR CNRS 7252, 123 Avenue A. Thomas, 87060 Limoges, France*
[2] *Université de Limoges, IRCER, UMR CNRS 7215, 12 Rue Atlantis 87068 Limoges, France*
[3] *SOLEIL Synchrotron, Gif-sur-Yvette, France*
[4] *iMATUS, University of Santiago de Compostela, Praza do Obradoiro S/N, Santiago de Compostela, Coruña 15782, Spain.*

vincent.couderc@xlim.fr

**Supplementary Material Summary**
This supplementary material provides detailed experimental and numerical results supporting the main manuscript. It includes analyses of second-harmonic generation (SHG) in multimode fibers, the dynamics of geometric parametric instability (GPI), evidence of refractive index changes, fiber characteristics, photothermal infrared spectroscopy, numerical simulations, and measurements of the nonlinear refractive index before and after the optical poling process.

**SM1 – Second-harmonic (SH) generation in a graded-index (GRIN) multimode 50/125 µm optical fiber before and after the poling process**
Section SM1 presents images of SHG recorded in a GRIN multimode optical fiber (Fiber 1 in the main manuscript) before and after the poling process. These recordings highlight the impact of doping in the fiber core, which induces local generation of non-phase-matched SH light. Additionally, the images illustrate the modifications introduced by optical poling on the transverse profile of the fiber core, as observed using a multiphoton (SH) microscope.

**SM2 – Evolution of geometric parametric instability (GPI) dynamics versus poling time**
Section SM2 demonstrates the enhancement of nonlinear GPI lines during the optical poling process. It also introduces the possibility of changes in both linear and nonlinear refractive indices within the multimode fiber core as a function of writing time, driven by the beating of fundamental (FF) and SH waves.

**SM3 – Evidence of linear refractive index changes after the optical poling process**
Section SM3 provides evidence of linear variations in the refractive index profile of the fiber core, resulting from multimode beating between SH and FF waves. These results suggest a potential molecular evolution of the silica structure during the poling process, extending beyond the well-documented charge displacement mechanisms.

**SM4 – Fiber characteristics and poling efficiency in three multimode optical fibers**
Section SM4 outlines the chemical and geometric characteristics of three multimode GRIN optical fibers subjected to the optical poling process. The primary finding is a significant difference in SHG efficiency between two fibers with similar chemical compositions. This observation suggests that the molecular characteristics of the fiber core may play a central role in determining poling efficiency.

**SM5 – Photothermal infrared (O-PTIR) spectroscopy of unpoled fiber cores**
Section SM5 presents the results of O-PTIR spectroscopy conducted on three unpoled multimode optical fibers. These analyses eliminate the influence of fictive temperature on the optical poling process and support the hypothesis that the molecular constitution of the fibers significantly contributes to the poling effect.

**SM6 – Numerical simulations showing additional self-cleaning process under poling effect**
Section SM6 presents results from numerical simulations demonstrating energy transfer toward the fundamental mode due to additional changes in the linear refractive index under the poling effect.

**SM7 – Measurement of the nonlinear refractive index before and after the poling process**
Section SM7 details measurements of the nonlinear refractive index of Fiber 1, before and after the poling process, using a novel method based on multiplexed coherent anti-Stokes Raman scattering (M-CARS). The results reveal a modification of the initial parabolic nonlinear profile following the inscription of $\chi^{(2)}$ nonlinearity.

**SM1 – SH generation in a GRIN multimode 50/125 µm optical fiber before and after poling process.**
Without poling, the cross-section image of the 50/125 µm germanium-doped multimode optical fiber exhibits non-zero SHG across the entire fiber core. This image, acquired using a multiphoton microscope at 800 nm in the femtosecond regime (Figure SM1(a)), reveals that doping the fiber core introduces local inhomogeneities in the silica structure. These inhomogeneities create non-centrosymmetric defects, leading to non-phase-matched local SHG. The transverse profile of this parametric fluorescence displays a quasi-parabolic distribution spanning the entire core area, closely following the doping concentration profile (Figure SM1(b)).
We reproduced this imaging after the optical poling process, using the poling method described in the main paper. Following more than 215 minutes of exposure, we remeasured the transverse image of the fiber's nonlinear response at approximately 20 centimeters from the entrance. The results show a significant increase in SHG intensity at the core center (Figure SM1(c)), presenting a quasi-circular imprint with a diameter of 15 µm. However, this image, measured on a 10 mm fiber segment, integrates the signatures of multiple poling periods as well as several self-imaging periods.
The circular pattern observed in the cross-section explains the presence of two opposing regions, arising from charge depletion induced by the poling effect with the fundamental beam [1, 2]. The evolution of the initial linear polarization state, transitioning from linear to elliptical along the propagation, drives charge displacement in all transverse directions, thereby forming an imperfect circular structure. At this longitudinal observation position, the fundamental beam remains slightly multimodal, which may also account for the complex structure of the $\chi^{(2)}$ nonlinearity observed in the image. One of the questions this paper answers is: does the area marked by the poling contain recordable molecular changes?

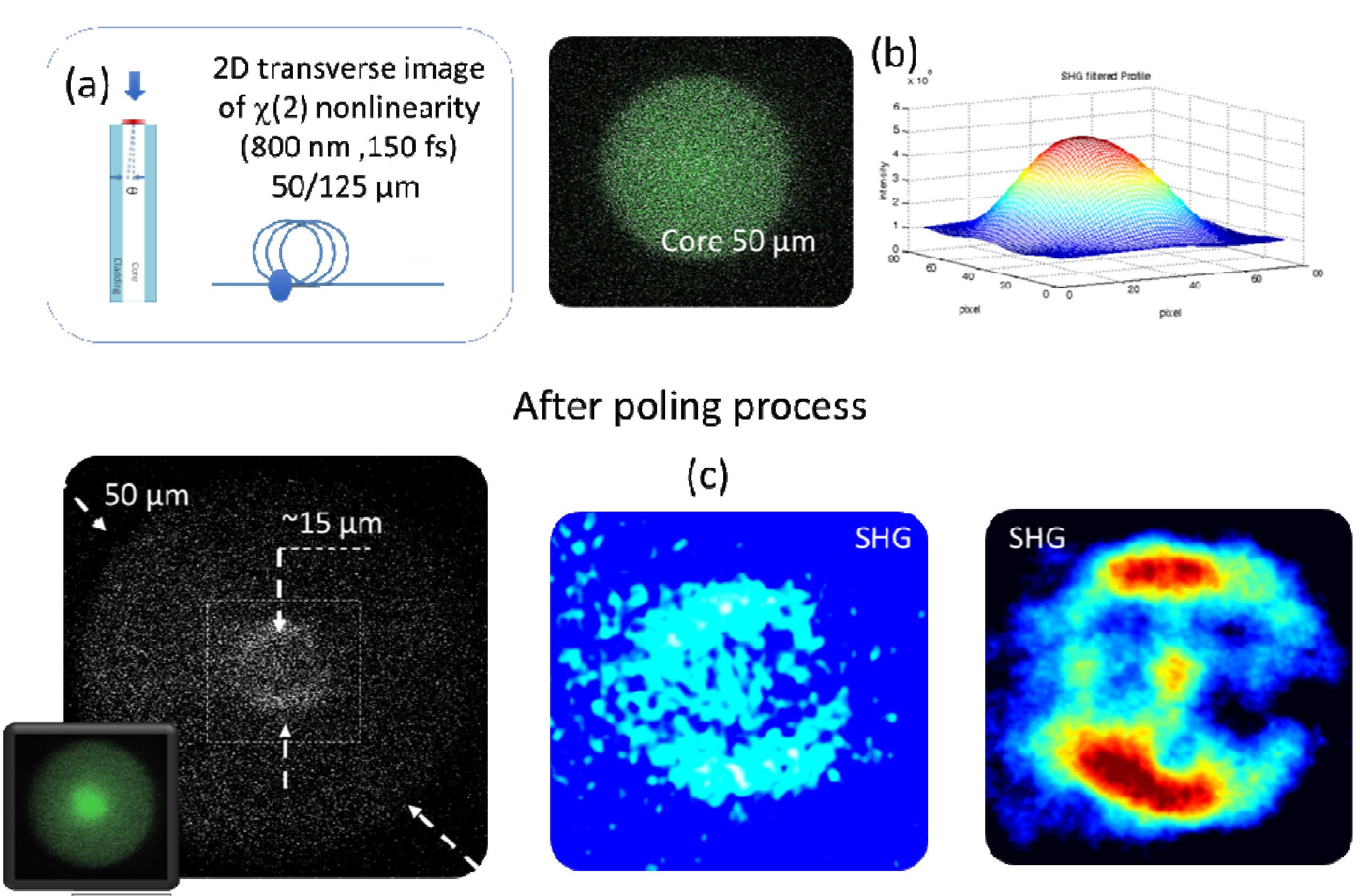

**Figure SM1:** Image of the cross-section of the 50/125 µm multimode fiber (Fiber 1) before and after the poling process. (a) Schematic of the experiment performed with a femtosecond laser source through a multiphoton microscope. (b) 2D and 3D images of the SHG obtained on 1 cm of fiber length before the poling process. (c) 2D images obtained after the poling process (215 min of exposure) with different colorimetry processes in order to better account for the transverse profile of the nonlinearity inscribed in the fiber after 20 cm of propagation.

**SM2 - Geometric Parametric Instability dynamics change versus poling time**

We recorded the output spectrum between 650 nm and 1200 nm as a function of the poling time (between 0 to 165 min) in a 50/125 µm GRIN multimode optical fiber (fiber 1). The FF input pulse had a duration of 640 ps for a maximum peak power of 52 kW. After 30 seconds of seeding with a type II KTP crystal, the second-order nonlinearity gradually increases and the SH beam reaches a maximum of 1.75 kW. The optical fiber used in this experiment is long enough (5 m) to excite the GPI sidebands [3]. We focused our attention on the first-order anti-Stokes line producing radiation near 735 nm. Its reciprocal line is obtained at 1926 nm, where the silica absorption begins to limit the energy transfer toward the parametric mixing. The longer the poling time, the greater the energy transfer to the sideband, which is not the case without a poling process. At 165 min, the maximum is reached for the 735 nm line while the fundamental infrared band is slightly depleted, also giving a little more energy to the Raman lines, which then appear as a limitation of the instability process efficiency.

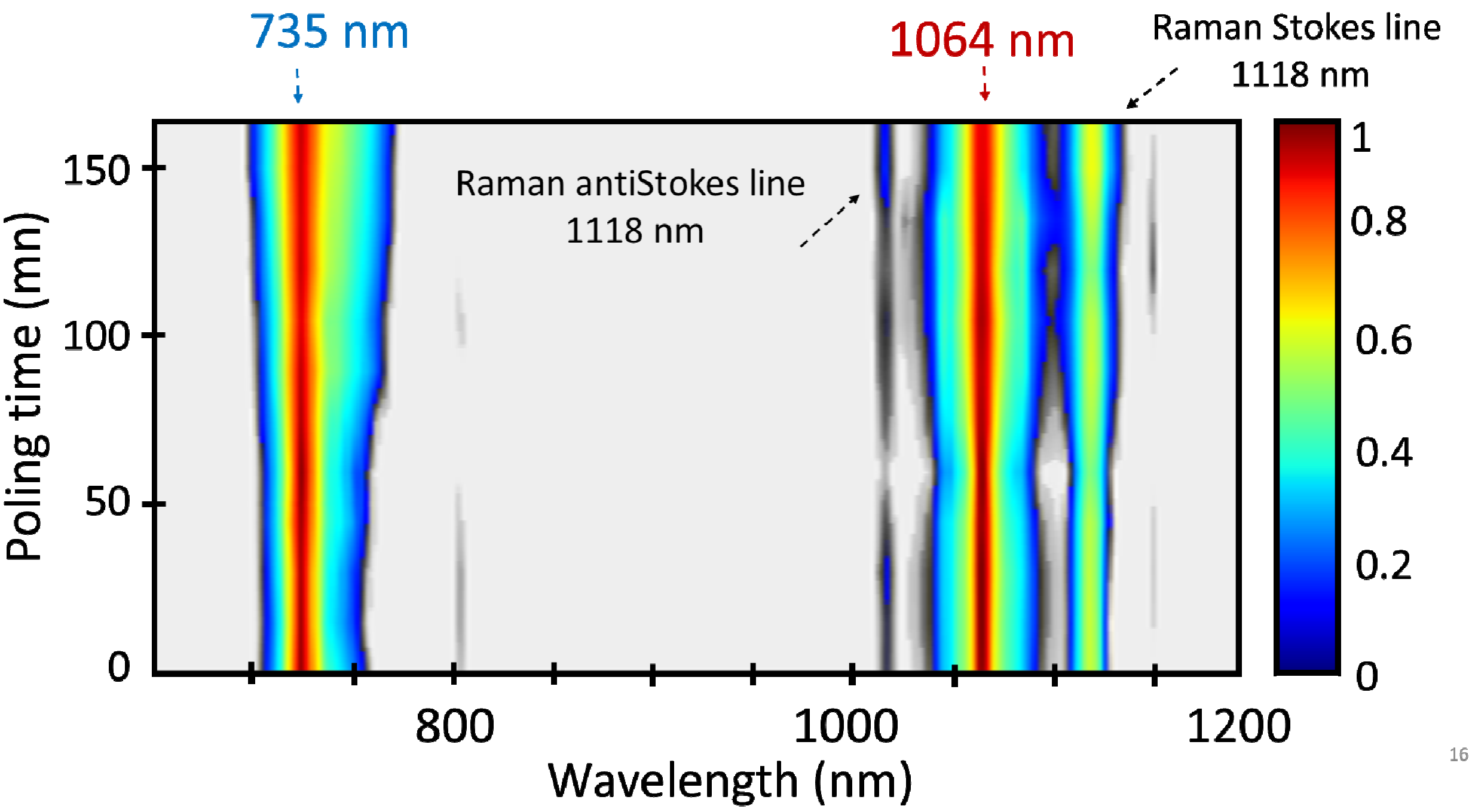

**Figure SM2:** Image of output spectrum versus the poling time. Optical fiber: 50/125 µm GRIN multimode fiber. Pump peak power: 52 kW. Seeding process with SHG: 30 s (7 kW for the SH, 45kW for the FF beam).

**SM3 - Evidence of the refractive index change after optical poling process**

Due to the evolution of nonlinear processes governed by the third-order nonlinear susceptibility during optical poling, we investigated the variation in the linear refractive index induced by the coherent beating between multimode FF and SH beams. This analysis was conducted using the Arden Photonics nPA600 profiler, which enables the measurement of the linear refractive index of the fiber.

Our observations revealed a slight, non-symmetric variation in the refractive index profile between the poled and unpoled cases. The most significant variation occurs at the beginning of propagation (5 cm), while this modification gradually diminishes after tens of centimetres (see Figures SM3(b) and SM3(c)).

We also observed a spatial inscription of the second-order nonlinearity in a zigzag pattern by using nonlinear microscopy process based on SHG (see Figure SM3(a)), which may explain the non-symmetric variation of the index profile after poling when measured with Arden Photonics nPA600. The zig-zag phenomenon is attributed to the spatial walk-off experienced by the FF and SH waves in the KTP crystal used for the initial seeding of the $\chi^{(2)}$ writing process. Indeed, the spatial delocalization of the SH relative to the pump beam in the birefringent crystal is transformed into a difference in orientation during the coupling in the multimode fiber through the focusing lens. In our case the maximum change in the linear refractive index is estimated to $6\ 10^{-4}$.

It is also worth noting that a change in the core refractive index of a single-mode fiber was previously suggested by Österberg et al. [4], who observed a variation in the transverse mode profile at the second harmonic. Similarly, Wasyhun Asefa Gemechu et al. [5] reported a slight variation in the linear refractive index following an optical poling effect in a transverse single-mode fiber. To date, no similar observations have been reported in multimode fibers. This change can be explained by the structural evolution of the poled fiber (see discussion in the main paper).

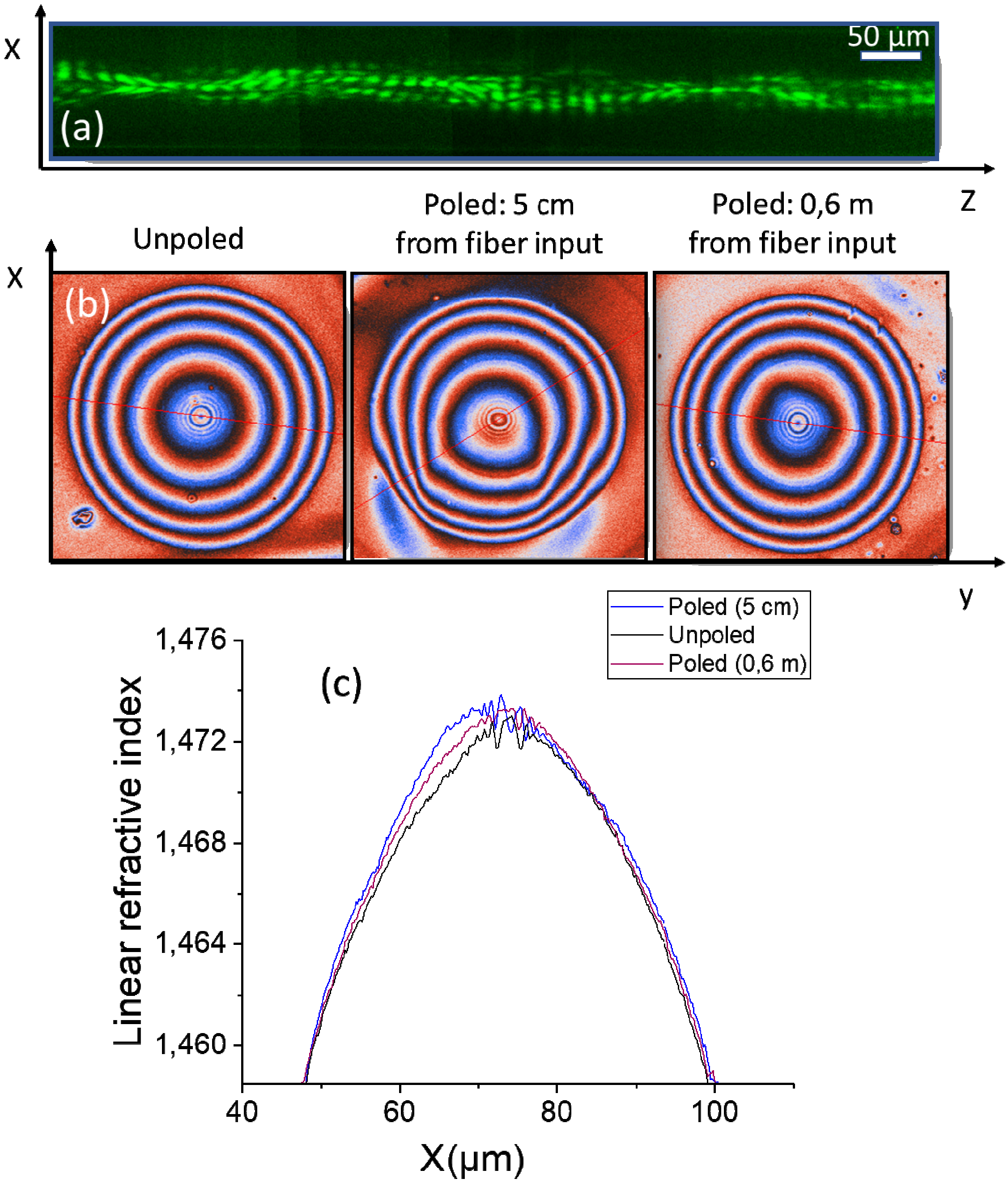

**Figure SM3:** Observation of the refractive index change in a 50/125 µm multimode optical fiber after the optical poling process. Peak pump power: 52 kW; Seeding process with SHG: 30 s (7 kW for the SH beam, 45 kW for the FF beam). (a) Image of the second-order nonlinearity (zig-zag) obtained by nonlinear microscopy imaging. The image illustrates the non-collinear propagation due to the beating between FF and SH beams. (b) Measurements of the transverse linear refractive index for two positions in the poled fiber (5 cm and 0.6 m from the input fiber). (c) Linear refractive index variation in the fiber core (Extracted from the 2D images).

**SM4 - Fiber features and poling efficiency in three multimode optical fibers**

We studied the optical poling process in three germanium-doped GRIN multimode fibers, two of them are geometrically and chemically identical. The third fiber, however, has a larger core diameter and a different doping concentration. The characteristics of these fibers are as follows:

- Fiber 1: 50/125 µm, GRIN Ge-doped silica fiber, Δn = 0.015 (supplied by Alcatel)
- Fiber 2: 50/125 µm, GRIN Ge-doped silica fiber, Δn = 0.015 (supplied by Thorlabs)
- Fiber 3: 62.5/125 µm, GRIN Ge-doped silica fiber, Δn = 0.028 (supplied by Thorlabs)

The doping concentration measurements and spatial dimensions of the fibers are shown in Figures SM4(a) and SM4(b).

We then performed the poling process on the three fibers using identical input parameters (peak power, pulse duration, repetition rate, and fiber length). Despite these controlled conditions, we observed significantly different efficiencies among the three fibers. While the difference in efficiency between Fibers 1 and 3 is somewhat expected due to their distinct characteristics, the disparity between Fibers 1 and 2, which are identical in terms of geometry and chemical composition, but manufactured by different suppliers, is far less intuitive.

The difference in efficiency is particularly striking: Fiber 1 achieved a SH peak power of 1.75 kW, whereas Fiber 2 exhibited more than 170 times less SH output, with a peak power of only 10.3 W.

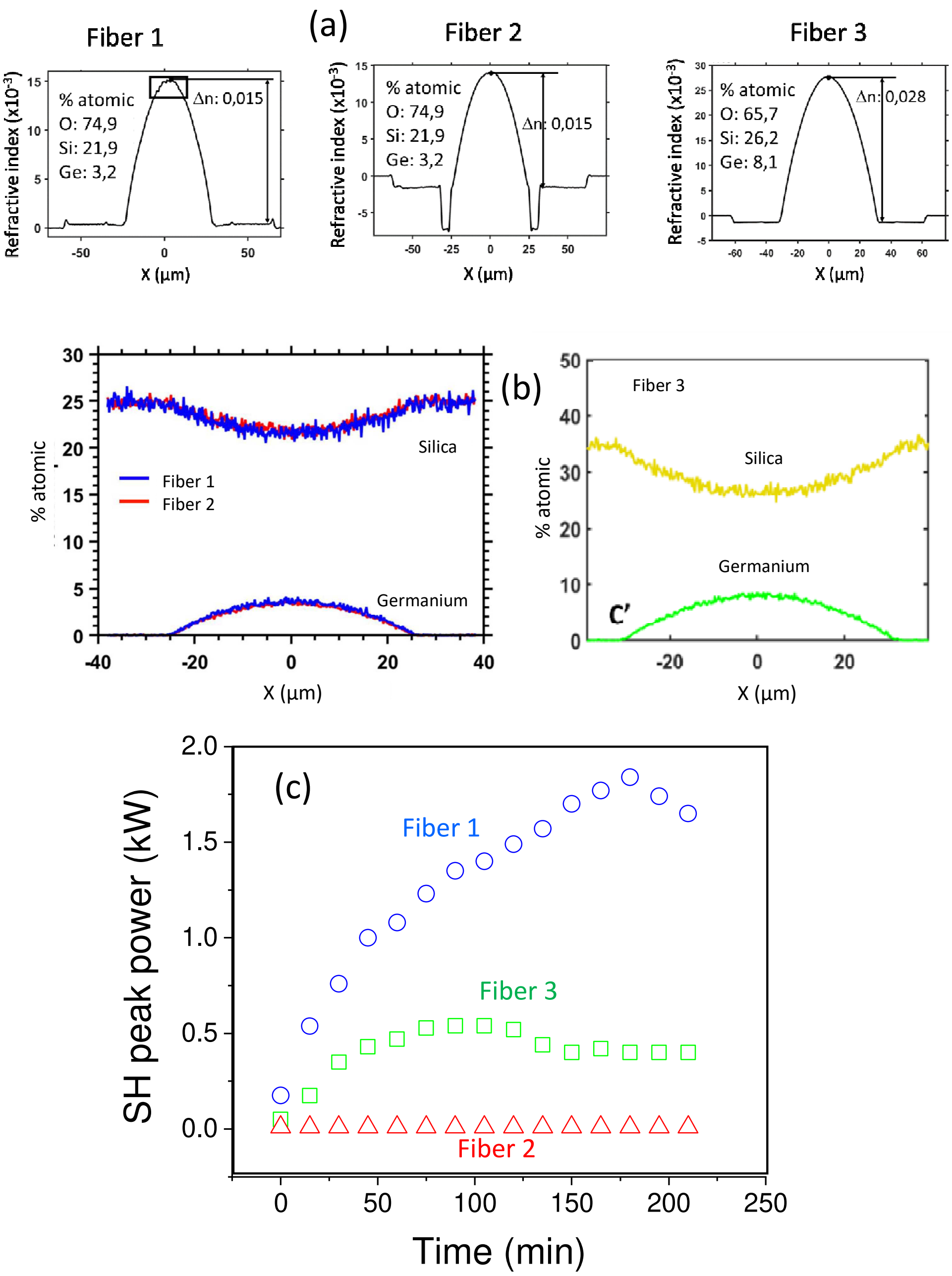

**Figure SM4:** Characteristics of the three multimode optical fibers tested for poling process and efficiency of the poling. (a) Refractive index profile and chemical constitution of the three multimode optical fiber: fiber

1: GRIN 50/125 µm from Alcatel; fiber 2: GRIN 50/125 µm from Thorlabs; Fiber 3: Grin 62.5/125 µm from Thorlabs. (b) % atomic constitution and profiles. (C) SHG efficiency for the three optical multimode fibers. FF peak power: 52 kW; FF pulse duration: 640 ps; Repetition rate 30 kHz.

**SM5 - Photothermal Infrared spectroscopy (O-PTIR) images of unpoled fiber cores**

The results of the principal component analysis of the concatenated data of all fibers, recorded with the O-PTIR spectroscope, are presented in Figure SM5(a). The score map for both components PC1 and PC2 highlights the lack of discrimination between fiber 1 and fiber 2. Only fiber 3, where the germanium concentration is the highest, can be discriminated with this method. Increasing the germanium concentration induces a decrease of the mode at 1120 $cm^{-1}$, which is clearly highlighted by the visualization of the score map (Figure SM5(b)) and the PC1 loading reflecting the shift of the peak to 1120 $cm^{-1}$. The similarity of the data between fibers 1 and 2 excludes the impact of the fictive temperature during fiber drawing and therefore cannot explain the strong difference during the optical poling step. The violin plot in Figure SM5(c) of the projected data in PC1 illustrates this analysis. Component 2 (PC2) can be correlated with the variation in the half-maximum width of the mode, and therefore with the germanium concentration. For better resolution of variations linked to the germanium concentration profile and $T_F$ distribution, the O-PTIR data were truncated around the peak at 1120 $cm^{-1}$ ([1061-1194]).

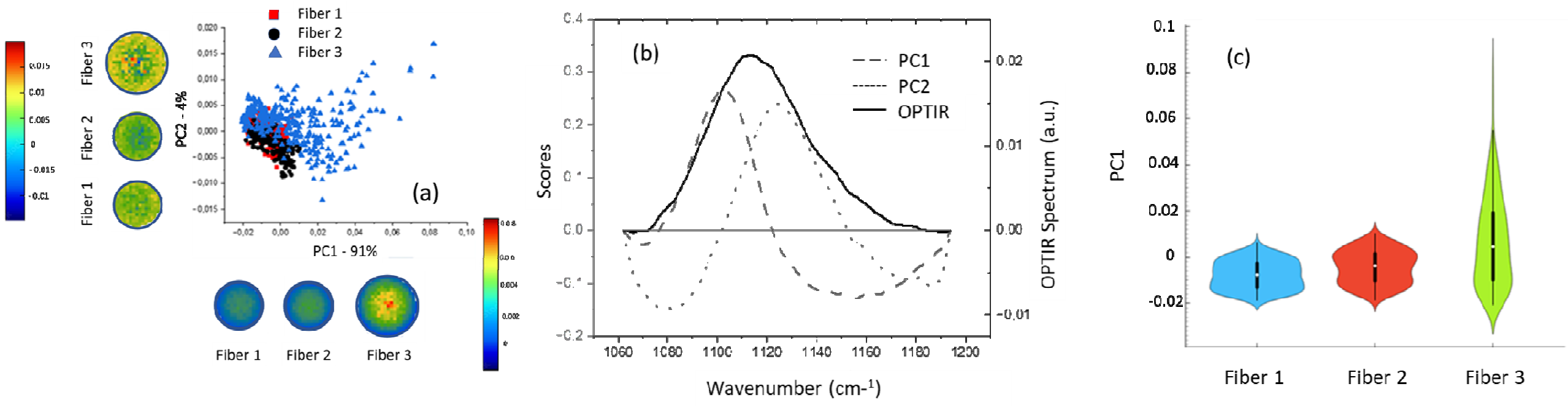

**Figure SM5:** (a) Component 1 (PC1) and 2 (PC2) of the scores map for the three fibers (Fiber 1: 50/125 µm Alcatel; Fiber 2: 50/125 µm Thorlabs; Fiber 3: 62.5/125 µm Thorlabs) and O-PTIR images of fiber cores reconstructed on the weight of components PC1 and PC2 respectively. (b): PC1 and PC2 loadings associated to a typical O-PTIR spectrum for visual support. (c) Violin representation of the data projected in PC1.

**SM6-Numerical simulation showing additional self-cleaning process under poling effect**

It is well known that the optical poling process modifies the electronic equilibrium within the material due to the creation of a static electric field and photocurrents, thereby locally altering the linear refractive index, as we have measured in multimode GRIN optical fibers (see Figure SM3). This process of writing $\chi^{(2)}$ nonlinearity also affects the molecular structure of the fiber core, forcing the large rings of 5-6 $SiO_4$ tetrahedra to decay into smaller rings composed of 4-3-2 tetrahedra. This restructuring modifies both the linear and nonlinear refractive indices within the material.

Here, we have developed (see "Methods" in the main paper) a numerical model demonstrating that local deformations of the liner refractive index can alter the energy distribution among modes, particularly modifying the energy propagated by the fundamental mode. We developed a first numerical model based on local and evolving refractive index variations along the propagation path. Initially, the refractive index distribution is irregular in the spatial domain at the beginning of the fiber, mimicking the multimode poling process, whereas it becomes more regular under the self-cleaning process (see Figure 1 of the main paper). The maximum change in the refractive index is similar to that measured on poled fibers (~6 x $10^{-4}$ at maximum, see Figure SM3).

Under these conditions, we plotted in Figure SM6 the power propagating in the fundamental mode with and without the spatial distortion introduced by the poling process. With the local refractive index change, the energy in the fundamental mode can be improved by between 5% and 20%, systematically demonstrating a booster effect introduced by poling.

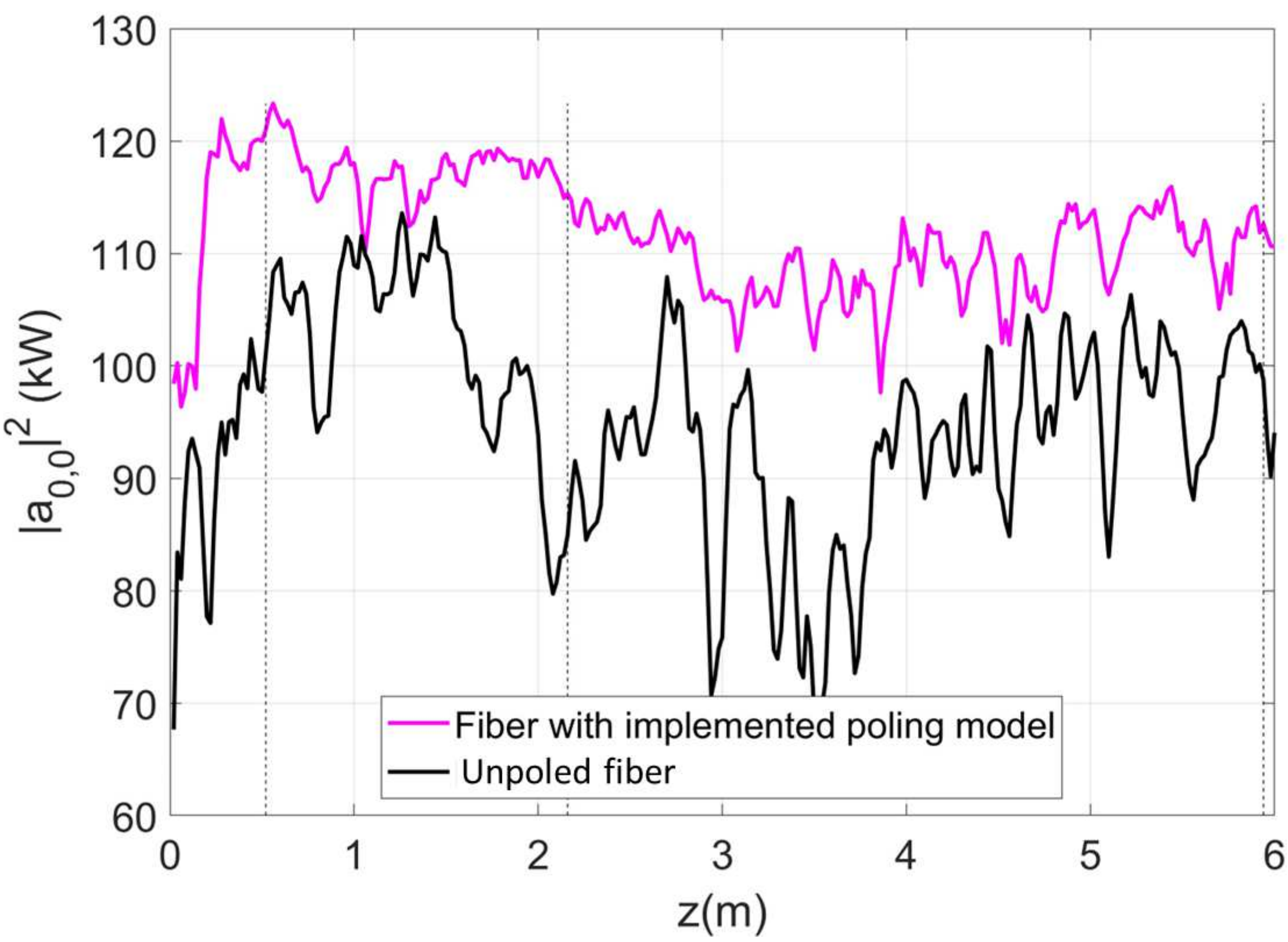

**Figure SM6:** Energy propagating on the fundamental mode of the 50/125 µm optical fiber 1, with (Pink) and without (Dark) poling process.

**SM7 - Nonlinear refractive index measurement before and after poling process**

Because of the enhancement of the Kerr spatial self-cleaning and the GPI spectral band generation, both of them relying on the linear refractive index phase matching and the third-order nonlinearity, we measured the nonlinear refractive index of unpoled and poled fibers. Our goal was to evaluate the impact of the poling process on the third-order nonlinear susceptibility. This measurement was performed across the entire transverse profile of the fiber core using a recent method based on multiplex coherent anti-Stokes Raman scattering (M-CARS) spectroscopy [6, 7]. This technique allowed us to extract the non-resonant background of the M-CARS process, thereby revealing the nonlinear refractive index of the doped fiber. Compared to the Z-scan technique [8], which requires a longitudinal scan for each spatial measurement, the M-CARS method is significantly faster and thus may produce 2D images.

Our objective was to observe differences between the poled and unpoled multimode GRIN fibers. The spectral measurement ranged from 650 $cm^{-1}$ to 2800 $cm^{-1}$ and has been shortened between 1850 $cm^{-1}$ and 2800 cm$^{-1}$ in order to avoid any vibrational contribution mainly appearing between 0 to 1200 cm$^{-1}$ in silica. Thus, only the contribution of the non-resonant background may be observed (Pure Kerr contribution).
The first result concerns the parabolic evolution of the unpoled fiber1 profile, which clearly shows a direct dependence of the Kerr nonlinearity versus Ge doping. This demonstrates that the nonlinear parabolic refractive index can be extracted using our M-CARS measurement (Figure SM7(a)).
Subsequently, we also measured the poled optical fiber 1 (figure SM7(b)). We observe a slight difference between the two maps. After normalization, we find that the core of the poled fiber exhibits significant distortions reducing the central part of the curve where poling has been mainly identified. This observation suggests a limited, but noticeable, impact of the poling process on the pure susceptibility $\chi(3)$, It is also important to note that the accuracy of our measurements does not allow us to precisely assess the absolute change in Kerr response in the poled fiber.

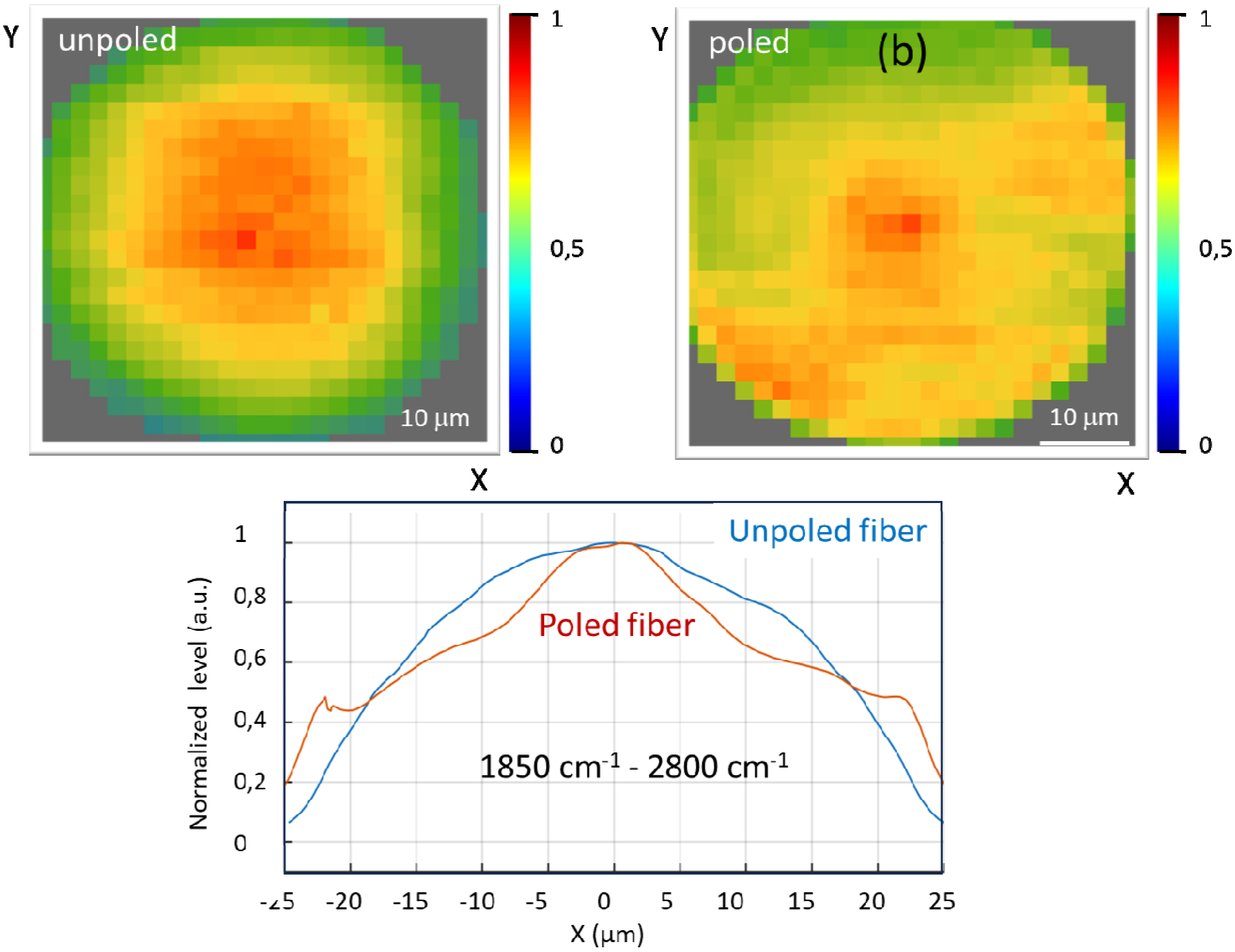

**Figure SM6:** M-CARS image of the poled and unpoled fibers 1 for wavenumbers between 1850 cm$^{-1}$ and 2800 cm$^{-1}$. (a) 2D transverse profile of the pure Kerr nonlinearity of the unpoled fiber 1. (b) 2D transverse image of the nonlinear refractive index of the poled fiber 1. (c) 1D profile of the poled and unpoled fiber 1.